\documentclass[twocolumn,tighten]{aastex631}
\usepackage{orcidlink}
\hypersetup{colorlinks, citecolor=blue, linkcolor=blue}
\definecolor{fgreen}{rgb}{1,0.6,0.1}

\newcommand{\mne}[1]{$-$}

\newcommand{\qdist}[1]{\ifmmode\langle#1\rangle\else\textlangle#1\textrangle\fi}

\newcommand{\kms}{km\,s$^{-1}$}

\newcommand{\hi}{\mbox{H\textsc{i}}}
\usepackage{soul}
\setstcolor{blue}

\received{\today}
\revised{\today}
\accepted{\today}
\submitjournal{ApJ Letters}

\shorttitle{Interacting early-type dwarf galaxies}
\shortauthors{Paudel et al.}

\begin{document}

\title{Rejuvenating Star Formation Activity in an Early-type Dwarf Galaxy, LEDA\,1915372, with Accreted HI Gas}

\author[0000-0003-2922-6866]{Sanjaya Paudel~\orcidlink{0000-0003-2922-6866}}
\affiliation{Department of Astronomy, Yonsei University, Seoul, 03722, Republic of Korea}
\affiliation{Center for Galaxy Evolution Research, Yonsei University, Seoul, 03722, Republic of Korea}

\author[0000-0002-1842-4325]{Suk-Jin Yoon~\orcidlink{0000-0002-1842-4325}}
\affiliation{Department of Astronomy, Yonsei University, Seoul, 03722, Republic of Korea}
\affiliation{Center for Galaxy Evolution Research, Yonsei University, Seoul, 03722, Republic of Korea}

\author[0000-0003-2722-8841]{Omkar Bait~\orcidlink{0000-0003-2722-8841}}
\affil{Observatoire de Gen\`eve, Universit\'e de Gen\`eve, 51 Chemin Pegasi, 1290 Versoix, Switzerland}

\author{Chandreyee  Sengupta}
\affiliation{Purple Mountain Observatory, Chinese Academy of Sciences, Nanjing, 210034, People’s Republic of China}

\author[0000-0003-0960-687X]{Woong-Bae G. Zee~\orcidlink{0000-0003-0960-687X}}
\affiliation{Department of Astronomy, Yonsei University, Seoul, 03722, Republic of Korea}
\affiliation{Center for Galaxy Evolution Research, Yonsei University, Seoul, 03722, Republic of Korea}

\author{Daya Nidhi Chhatkuli}
\affil{Central Department of Physics, Tribhuvan University, Kirtipur, Kathmandu, Nepal}

\author{Binod Adhikari}
\affil{Department of Physics, St. Xavier's College, Tribhuvan University, Kathmandu, Nepal}
\affil{Department of Physics, Patan Multiple College, Tribhuvan University, Nepal}

\author{Binil Aryal}
\affil{Central Department of Physics, Tribhuvan University, Kirtipur, Kathmandu, Nepal}

\correspondingauthor{Suk-Jin Yoon}
\email{sjyoon0691@yonsei.ac.kr }

\begin{abstract}
We report a rare astrophysical phenomenon, in which an early-type dwarf galaxy (dE), LEDA\,1915372, is accreting gas from a nearby star-forming dwarf galaxy, MRK\,0689, and is  rejuvenating star-formation activity at the center.
Both LEDA\,1915372 and MRK\,0689 have similar brightness of $M_{r}$ = $-$16.99 and $-$16.78 mag, respectively. 
They are located in a small group environment, separated by a sky-projected distance of 20.27 kpc (up to 70 kpc in three dimension), and have a relative line-of-sight radial velocity of 6 \kms.
The observation of 21 cm emission with the Giant Metrewave Radio Telescope provides strong evidence of interaction between the pair dwarf galaxies in terms of neutral hydrogen (\hi) morphology and kinematics. In particular, the \hi\ map reveals that the two galaxies are clearly connected by a gas bridge, and the gas components of both 
LEDA\,1915372 and MRK\,0689 share a common direction of rotation.
We also find that the \hi\ emission peak deviates from LEDA\,1915372 toward its optical blue plume, suggesting a tidal origin of ongoing central star formation. 
Our findings provide a new path to the formation of blue-cored dEs.
\end{abstract}

\keywords{Dwarf galaxies (416), Early-type galaxies (429),  Galaxy interactions (600), Galaxy pairs (610), Galaxy nuclei (609), \hi\ line emission (690) }

\section{Introduction}

The most common type of galaxy found in dense environments such as galaxy groups or clusters are early-type dwarf galaxies \citep[dEs;][]{Binggeli87,Binggeli90,Ferguson94}. 
These galaxies are typically red in color and have a smooth and regular appearance. 
However, recent studies have shown that a significant fraction of dEs display substructural features such as spiral arms, bars, and central nuclei \citep{Lisker06,Lisker07}. 
In addition to the structural diversity, variations in color suggest the presence of a range of stellar populations within dEs \citep{Paudel10,Paudel11}.

A particularly intriguing subgroup of dEs is the one that contains young central stellar populations, known as blue-centered dEs \citep{Lisker06b,Urich17,Hamraz19}. 
They have been observed in a variety of environments, from completely isolated regions to galaxy groups and clusters, and in some instances, they even constitute the majority of the dE population \citep{Pak14,Chung19,Rey23}. 
However, they are nearly absent in the massive Coma cluster \citep{Brok11}.

While classical dEs are typically not believed to have ongoing star formation or significant gas or dust content, there are exceptions such as NGC\,205 \citep{Richter87,Haas98}, NGC\,185 \citep{Marleau10} and IC\,225 \citep{Gu06} that have been observed to have such features. 
Additionally, \cite{Conselice03} reported a 15\% detection rate of neutral hydrogen (\hi) in dEs in the Virgo Cluster. 
The Arecibo Legacy Fast ALFA (ALFALFA) survey identified 12 dEs that possess \hi\ above the detection limit \citep{Hallenbeck12,Hallenbeck17,Haynes18}. 
The presence of \hi\ in these low-mass early-type galaxies raises questions about their origin.

\begin{figure*}
\includegraphics[width=18cm]{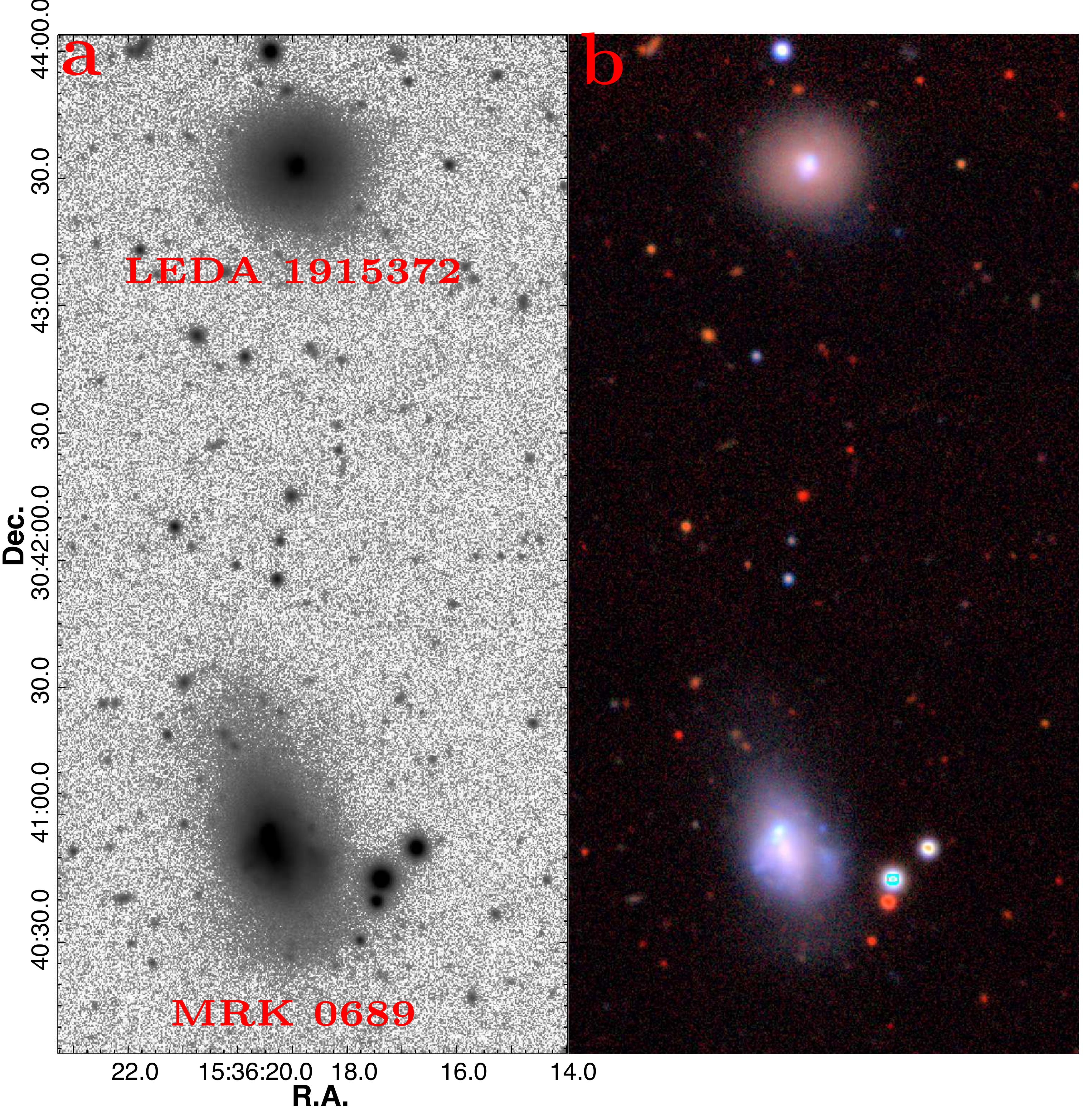}
\label{mainfig}
\caption{(a) The Legacy survey coadded $g$-$r$-$z$-band image of the system with a field of view of 1$\arcmin$$\times$2$\arcmin$. (b) $g$-$r$-$z$-band combined tricolor image.  }
\end{figure*}

Several scenarios have been proposed to explain the presence of cold gas and the central star-forming nature of early-type galaxies \citep{Chun20}.
These galaxies could be in a late transitional stage, where their internal star-forming gas has been stripped, and only small amounts of \hi\ are being replenished internally \citep{Koleva13}. 
Alternatively, the detected \hi\ could have recently been acquired from external sources, such as tidal interactions with nearby companions or the accretion of smaller dwarf galaxies.
A mixture of these different processes is also likely possible \citep{Oosterloo07,Serra08}.

In this work, we report a blue-centered dE, LEDA\,1915372, which appears to have acquired star-forming gas from its nearby companion galaxy, MRK\,0689. 
To perform a detailed analysis of the gas properties in this interacting system, we conduct radio observations using the Giant Metrewave Radio Telescope (GMRT) to map the 21 cm \hi\ emission in the region.

\section{Identification}

Our primary objective is to create an extensive catalog of dE galaxies located in diverse environments. 
To achieve this goal, we conducted a thorough visual search of color images from the SDSS and Legacy surveys within the local volume ($z$ $<$ 0.02), systematically identifying and inspecting relevant objects \citep{Paudel11}. 
In this Letter, we present our findings on LEDA\,1915372, a dE galaxy located in a less dense group environment. 
LEDA\,1915372 is a blue-centered dE with ongoing central star-forming activity.
It is similar to other central starburst dE galaxies that we have previously studied in \citet{Paudel20}, but it features a faint tidal tail and is located near another star-forming dwarf galaxy, MRK\,0689.

\begin{table}[h]
\caption{Physical properties of LEDA\,1915372  and  MRK\,0689}
\begin{tabular}{lcc}
\hline
Properties                     & LEDA\,1915372  &  MRK\,0689      \\
\hline          
R.A. (h:m:s)                   &  15:36:18.90   &  15:36:19.441   \\
decl. (d:m:s)                 &  30:43:32.88   &  30:40:56.34    \\
$z$                             &  0.005839      &  0.005859       \\
$D$ (Mpc)  & 26.45 &  26.52 \\
$M_{r}$ (mag)                  &  $-$16.99$\pm$0.01      &  $-$16.78$\pm$0.01       \\
$g-r$   (mag)                  &  0.63$\pm$0.01\tablenote{after masking the central region. Without masking the central region, the value of $g-r$ color is 0.45 mag.}          &  0.33$\pm$0.01           \\
SFR ($M_{\sun}$ yr$^{-1}$)     &  0.012$\pm$0.002         &  0.087$\pm$0.002        \\
12+Log(O/H)   &  8.15$\pm$0.05         &  8.18$\pm$0.05                        \\
Log($M_{*}/M_{\sun}$)         &  8.81$\pm$0.07           &  8.73$\pm$0.07           \\
Log($M_{HI}/M_{\sun}$)         &  9.09$\pm$0.01          &  9.16$\pm$0.01          \\
\hline
\end{tabular}
\label{ptab}
\end{table}

At the position R.A. = 15:36:18.90, decl. = +30:43:32.88, and with a redshift of $z$ = 0.0059, we discovered a rare dE, LEDA\,1915372, in close proximity to another star-forming dwarf galaxy, MRK\,0689. 
Both galaxies possess faint tidal tails, but MRK\,0689 appears to be more disturbed. Based on the assumption that they lie within the same sky plane, the sky-projected separation between the two galaxies is 20.27 kpc  (2$\arcmin$.62). However, considering their individual distances from us, the actual three-dimensional separation could extend up to 70 kpc. A relative line-of-sight velocity between them is 6 \kms. LEDA\,1915372 and MRK\,0689 have comparable brightness, with $M_{r}$ = $-$16.99 and $-$16.78 mag for LEDA\,1915372 and MRK\,0689, respectively. 
The physical properties of these galaxies are summarized in Table \ref{ptab}.

\begin{figure*}[h]
\includegraphics[width=18cm]{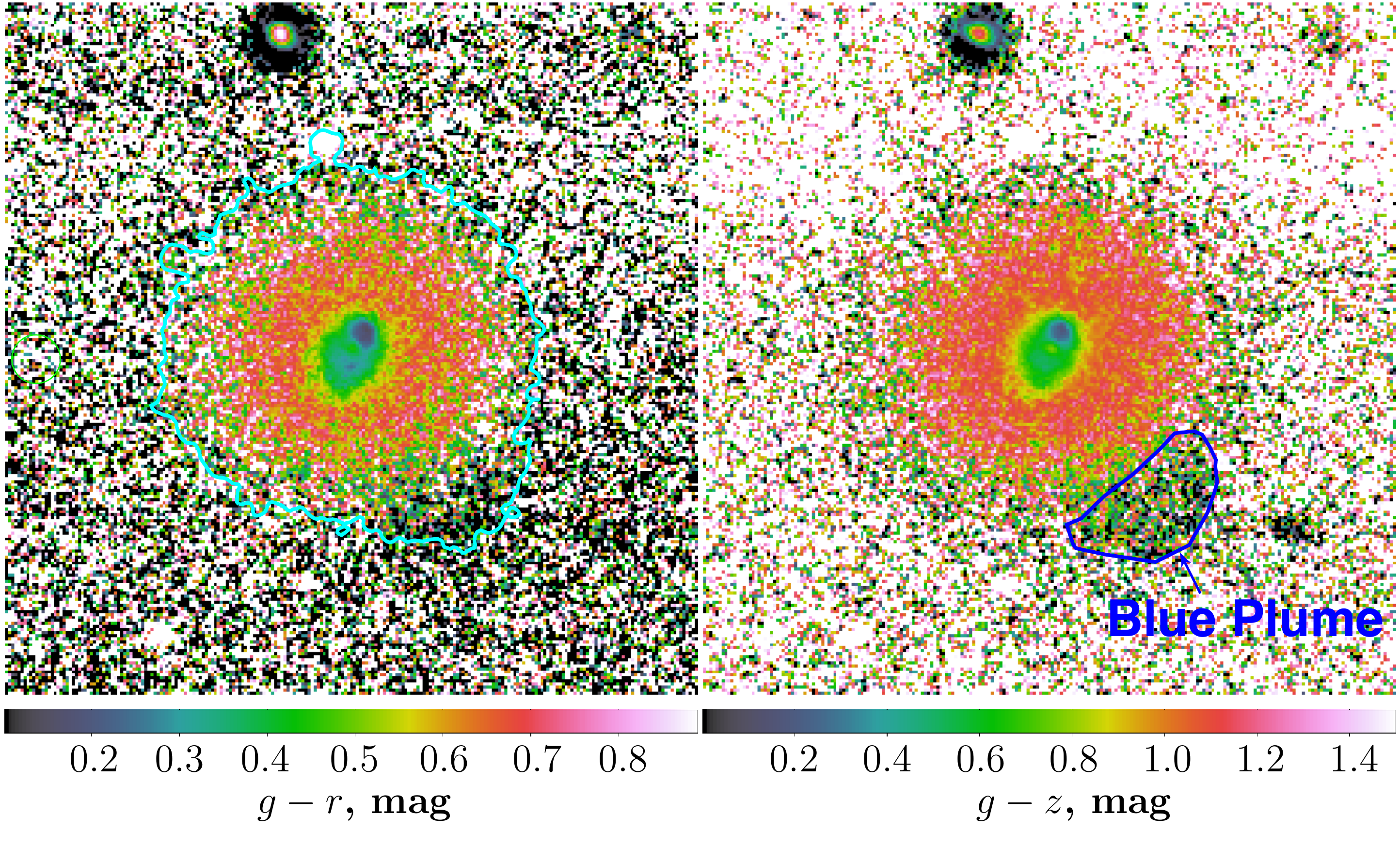}
\label{cmap}
\caption{In left and right, respectively, the $g-r$ and $g-z$ color maps of LEDA\,1915372, where a cyan contour represents a surface brightness level of 25.5 mag arcsec$^{-2}$ in $g$-band. An extended tidal tail named Blue Plume in LEDA\,1915372 is also marked. }
\end{figure*}

The pair is situated in a relatively sparsely populated region, with NGC\,5961 being their nearest bright galaxy companion. 
Within a sky-projected radius of 700 kpc and a relative line-of-sight radial velocity of $\pm$700 \kms\ with respect to NGC\,5961, we identified eight galaxies.
The absolute $B$-band magnitude of NGC\,5961 is $-$18.24 mag, suggesting that it is not a particularly luminous galaxy.
Based on the number of galaxies in its vicinity and their radial velocity distribution, NGC\,5961 appears to form a small group of galaxies \citep{Kourkchi17}. The projected physical separation between  NGC\,5961 and LEDA\,1915372 is 117 kpc, with a relative line-of-sight redial velocity of 163 \kms\ between the two galaxies.
For this work, the distances to the member galaxies are derived from the Numerical Action Methods model\footnote{http://edd.ifa.hawaii.edu/NAMcalculator/}  \citep{Shaya17,Kourkchi20}. The distance to the  LEDA\,1915372, MRK\,0689, and NGC\,5961 are 26.45, 26.52, and 27.45 Mpc, respectively.

\begin{figure}
\includegraphics[width=8.5cm]{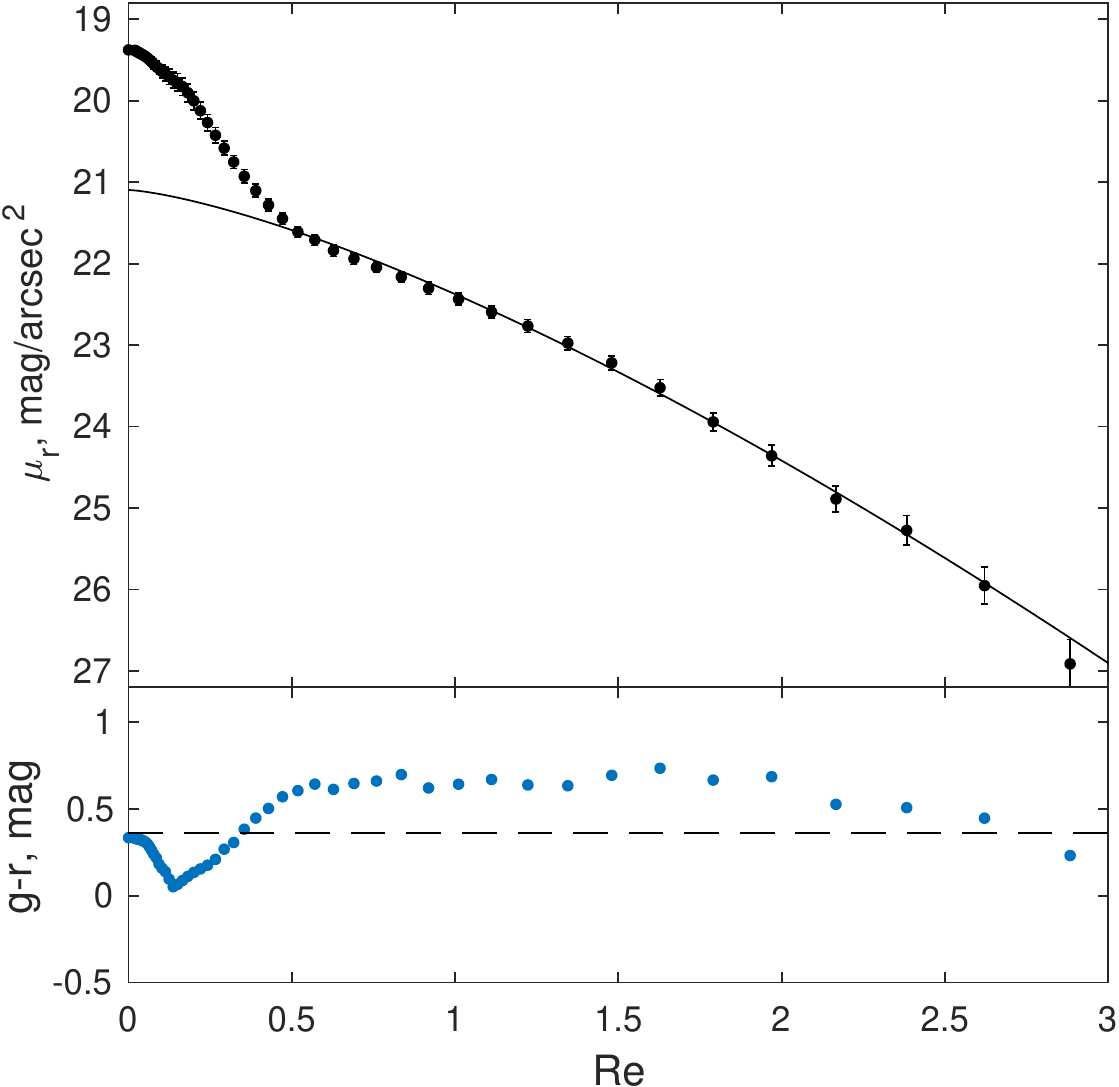}
\caption{Top: The $r$-band surface brightness profile of LEDA\,1915372. The solid line represents a best-fitted S\'ersic function. Bottom: a $g-r$ color profile along the major axis. The dashed horizontal line represents the $g-r$ color of the Blue Plume.}
\label{lightprof}
\end{figure}

 \begin{figure}
\includegraphics[width=8.5cm]{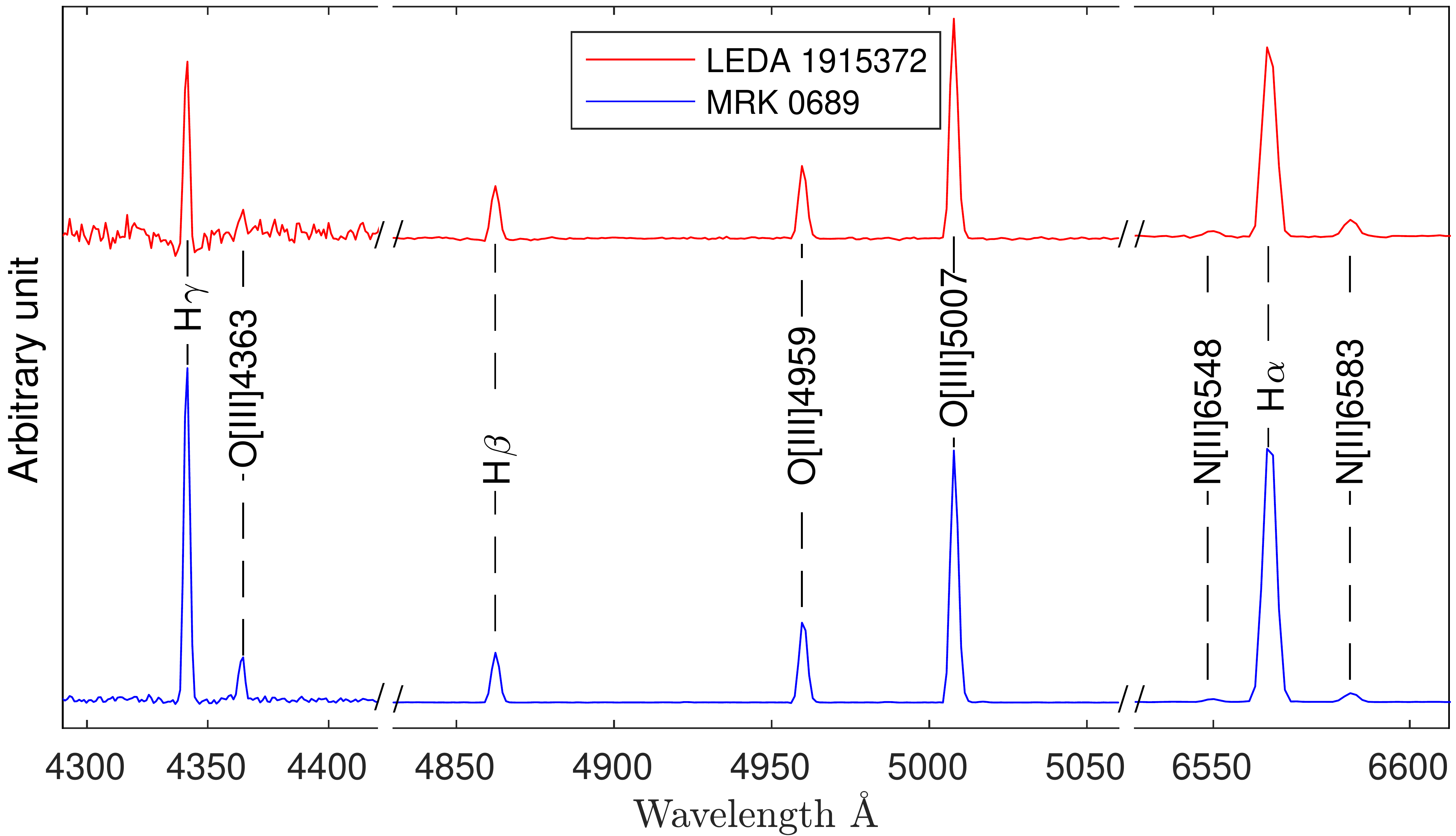}
\caption{SDSS optical spectra of  LEDA\,1915372 and MRK\,0689. The prominent emission lines that are analyzed to calculate the gas-phase metallicity are marked.   }
\label{spec}
\end{figure}

\begin{figure*}
\includegraphics[width=18cm]{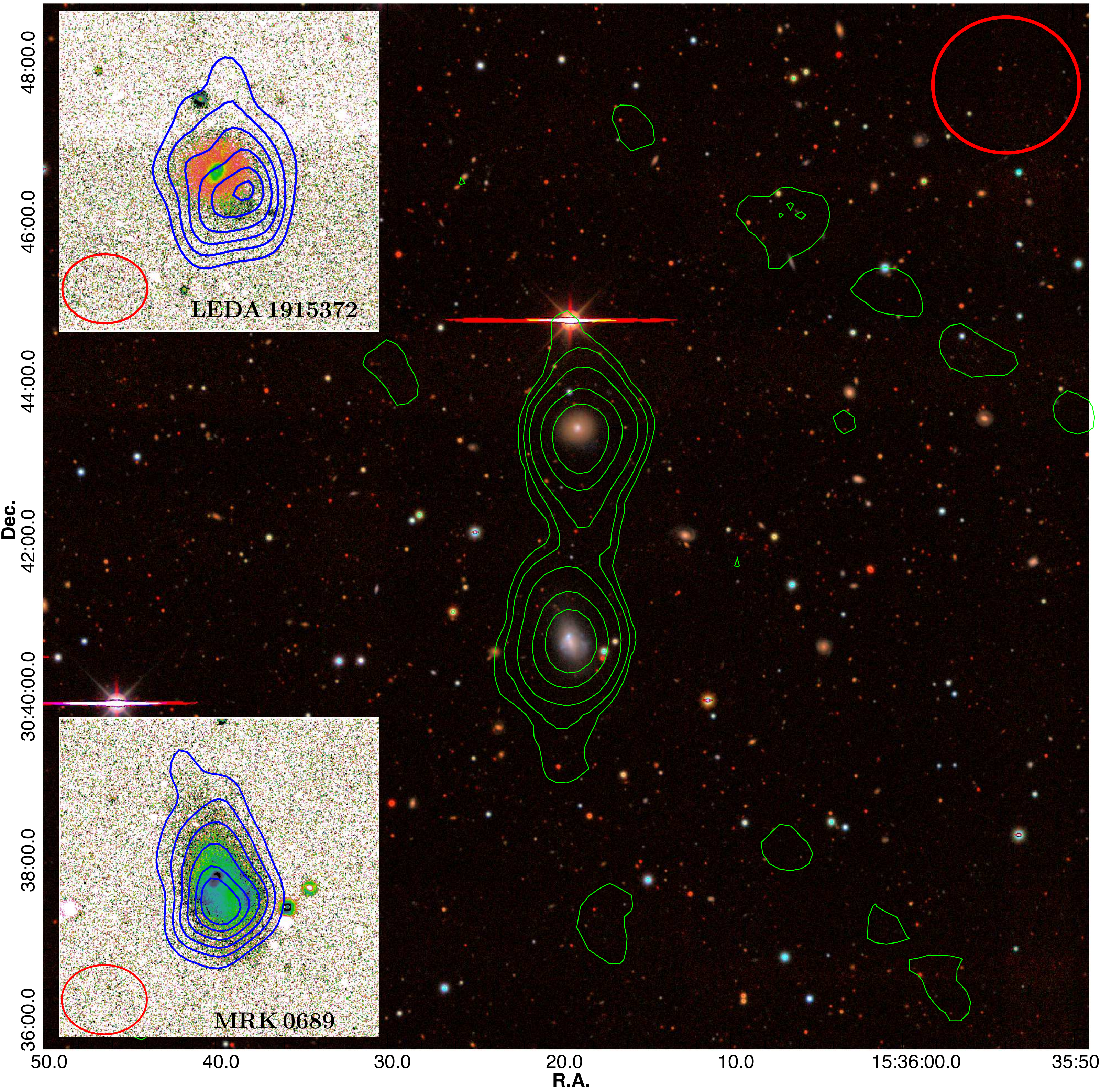}
\caption{ Integrated \hi\ contours from the GMRT low-resolution map overlaid on the Legacy Survey $g$-$r$-$z$ combined image. The field of view of the image is  12$\arcmin$$\times$12$\arcmin$. The \hi{} column density levels are N(\hi) = 10$^{20}$ $\times$ (0.5, 1, 2 4, 5, and 6) cm$^{-2}$. The inset shows a zoomed-in view of the two interacting galaxies. The overlaid \hi\ contour is obtained from a high-resolution map with a beam size of 15$\arcsec$\!.3$\times$11$\arcsec$\!.9 which is overlaid on the $g-z$ color map. The \hi\ column density levels are N(\hi) = 10$^{21}$ $\times$ (0.5, 1, 1.5, 2, 2.5, and 2.8) cm$^{-2}$. The red ellipse in each map represents the respective beam size.  }
\label{himap}
\end{figure*}

\section{Data analysis}\label{data}

Our study greatly benefited from the abundance of multiwavelength data that was easily accessible through public archives. 
This allowed us to conduct a comprehensive analysis of the morphology and stellar population characteristics of our sources. 
Specifically, we conducted a multiwavelength investigation of the system, using archival images from the Legacy survey \citep{Dey19} and the Spitzer Space Telescope, covering a wide wavelength range from optical to infrared.
In addition, we observed the system using the GMRT to capture the \hi\ 21 cm line.

\subsection{Analysis of archival data}\label{analysis}

In Figure \ref{mainfig}(a), we display the Legacy $g$-band image of the region around LEDA\,1915372 and its companion, with a field of view of 1$\arcmin$$\times$2$\arcmin$. In Figure \ref{mainfig}(b), we show a tricolor image obtained by combining $g$, $r$, and $z$-band images. 
Despite both galaxies in the pair appearing disturbed, we observe no connection between them in the optical images. 
LEDA\,1915372 exhibits a regular dE appearance, with a noticeable blue color at its center.
Notably, we find a blue star-forming blob located off-center with a separation of 2$\arcsec$ from the center. 
This feature is more easily seen in the color map presented in the following panels.

Upon a careful examination of both the grayscale and tricolor images of LEDA\,1915372, we have detected the presence of an extended low-surface-brightness feature in the lower right region, which may be a tidal tail.
In Figure \ref{cmap}, the extended low-surface-brightness region is bluer compared to the dE main body, which we refer to as the ``Blue Plume.''
LEDA\,1915372 has an overall $g-r$ color of 0.46 mag, and after masking the central blue region, $g-r$ for the main body becomes 0.63 mag.
The Blue Plume has a $g-r$ color of 0.35 mag, which is significantly bluer than the main body of the galaxy.

We also performed surface photometry on the optical images of LEDA\,1915372. 
We utilized the legacy $r$-band images and subtracted the sky backgrounds using the same method described in \cite{Paudel23}.
To extract the galaxy's major-axis light profile, we employed the IRAF ellipse task, following the manual masking of any non-related background and foreground objects. 
We held the center and position angle constant during the ellipse fit, allowing only the ellipticity to vary. 
The center of the galaxy was calculated using the IRAF task $imcntr$, and the position angle was determined via multiple iterative ellipse runs.
In order to avoid complexity during the modeling process, we excluded the inner star-forming region. We utilized the $\chi^{2}$ minimization approach to fit the observed galaxy light profile to a S\'ersic function.

In Figure \ref{lightprof}, we present the result of surface photometry on the optical images of LEDA\,1915372. In the upper panel, the S\'ersic function effectively models the major axis light profile of this object, with an effective radius of 12$\arcsec$ (1.2 kpc) and an index $n$ = 0.92.
These characteristics are commonly found in dEs located within the Virgo cluster.
In the lower panel, we also provide a $g-r$ color profile along the major-axis, which reveals a noticeable color gradient between the inner and outer regions of the dE. Notably, the bluest peak is not located at the center, indicating that the star-forming blob is off-center. The average $g-r$ color value (0.35 mag) of the Blue Plume (shown by the dashed line) is similar to that of LEDA\,1915372's center (0.33 mag).

The SDSS has targeted both galaxies for spectroscopic observation, and we obtained their optical spectra from the SDSS archives. 
These galaxies display typical emission lines found in H \textsc{ii} regions. Among other emission lines, we also detected the [O \textsc{iii}]$\lambda$4363 auroral line in both galaxies, as shown in Figure \ref{spec}. The [O \textsc{iii}]$\lambda$4363 emission line is highly sensitive to temperature and enables us to directly measure the gas phase metallicity by calculating the electron temperature. To obtain the value of 12 + log(O/H), we applied a method outlined by \cite{Montero17}. The resulting values for 12 + log(O/H) are 8.15$\pm$0.05 and 8.18$\pm$0.05 dex for  LEDA\,1915372 and MRK\,0689, respectively. While using the empirical O3N2 method provided by \cite{Marino13}, which analyzes a combination of the line ratios  H$_{\alpha}$/[N \textsc{ii}] and [O \textsc{iii}]/H$_{\beta}$, we obtained the values for 12 + log(O/H) are 8.17$\pm$0.05 and 8.12$\pm$0.05 dex for  LEDA\,1915372 and MRK\,0689, respectively.

To measure the stellar mass and star formation rate (SFR) of the galaxies, our study utilized the Spitzer Space Telescope infrared imaging data. The Infrared Array Camera (IRAC) images at 3.6 and 4.5-microns and Multi-band Imaging Photometer (MIPS) at  24 $\mu$m were retrieved from the IRSA\footnote{https://irsa.ipac.caltech.edu/Missions/spitzer.html} archive database. Due to their shallow depth compared to the Legacy optical images, the IRAC images are utilized only for photometry. After carefully removing any unrelated foreground and background objects, and we performed aperture photometry. To derive the stellar mass, we used a calibration provided by \cite{Eskew12}, which allowed us to convert the IRAC 3.6 and 4.5 $\mu$m fluxes to the stellar mass.  We then calculated the SFR from MIPS  24 $\mu$m flux using the empirical formula provided by \cite{Rieke09}. The SFRs of LEDA\,1915372 and MRK\,0689 derived from the MIPS 24 $\mu$m fluxs are  0.012$\pm$0.002 and  0.087$\pm$0.002 $M_{\sun}$ yr$^{-1}$, respectively.

\subsection{Radio 21-cm observation}\label{radio}

To obtain a detailed understanding of the \hi\ distribution in LEDA\,1915372 and its companion galaxy MRK\,0689, we conducted \hi\ interferometric observations using the GMRT on 2018 August 03. 
The total on-source time of the observation was 6.46 hr. 
The observations were carried out with a baseband bandwidth of 16.67 MHz, split into 512 channels, providing a velocity resolution of approximately 7 \kms. 
The GMRT primary beam at the $L$ band is 24$\arcmin$, and the images presented in this study are synthesized at two different resolutions of beam sizes of 16$\arcsec$$\times$13$\arcsec$ and 55$\arcsec$$\times$51$\arcsec$ for high and low resolutions, respectively.

\begin{figure*}[h]
\includegraphics[width=18cm]{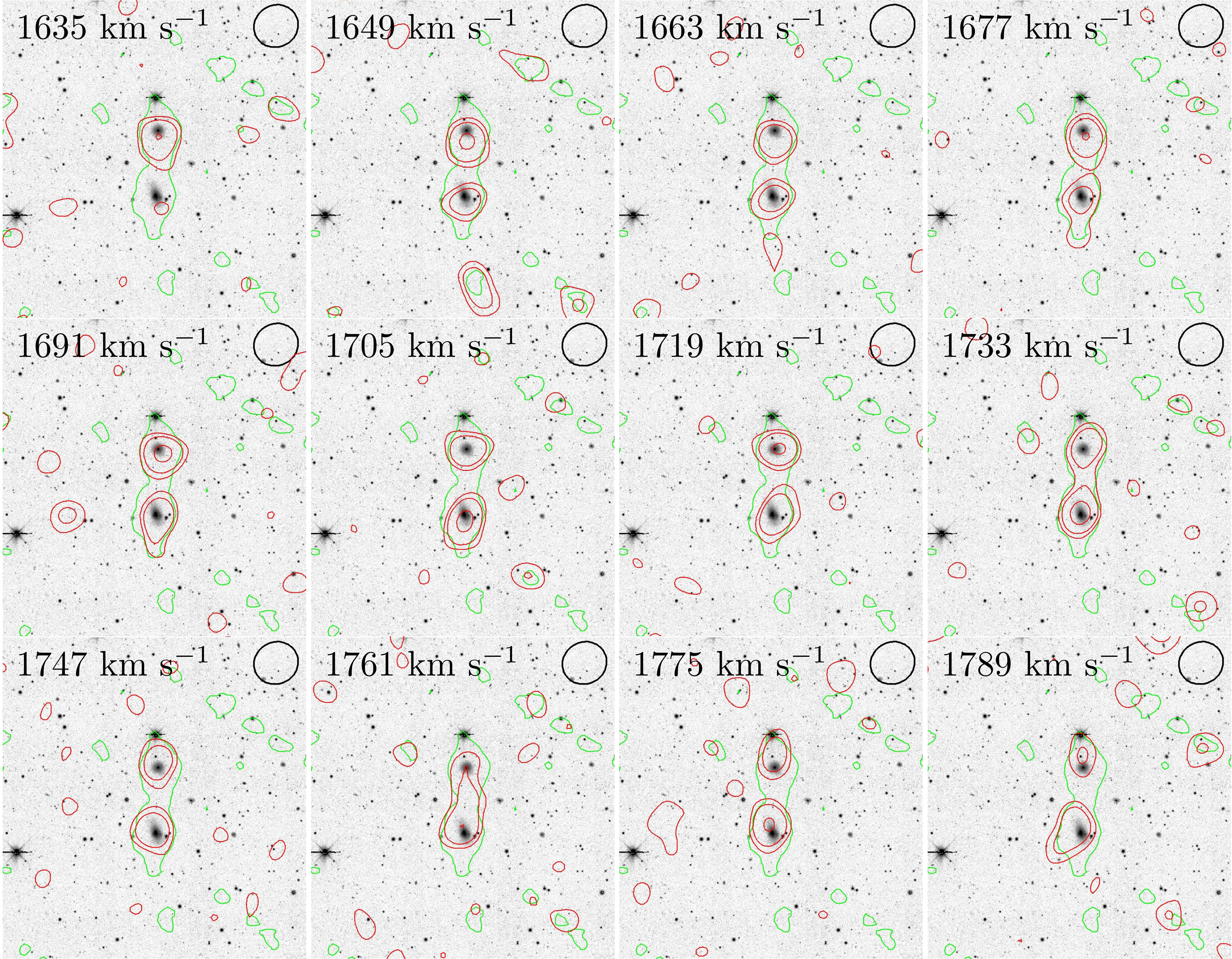}
\caption{\hi\,velocity channel contour map of the system.  The contours are shown at 1$\times$10$^{-3}$$\times$(2, 5) Jy beam$^{-1}$. To guide the position of the channel map, we also show a gray-scale image of the system in the background where the green contours represent the lowest column density level shown in Figure \ref{himap}.  The observed beam size is shown in the top right corner. }
\label{chmap}
\end{figure*}

The data were processed using the modular CASA v6.5.2 \citep{McMullin07,CTeam22}. 
The raw visibilities underwent several loops of initial flagging and calibration (amplitude, phase, and bandpass). 
Subsequently, the continuum was estimated in the line-free channels and subtracted using the CASA task ``uvcontsub".  
The continuum-free data were then imaged using “tclean” with a velocity resolution of 14 \kms\ at two different spatial resolutions of $\sim$55$\arcsec$$\times$51$\arcsec$ and $\sim$16$\arcsec$$\times$13$\arcsec$. We used a robust of 0.5, and a CLEAN threshold of 2$\sigma$ with a single CLEAN mask extending up to a primary-beam gain level of 0.2 (pbmask=0.2). The moment-0 and moment-1 maps were then constructed using the “immoments” task in CASA\footnote{The entire processing pipeline will be described in detail in Bait et al. (2023) in prep, and available online.}, by manually selecting the channels showing emission and a simple threshold blanking at 2$\sigma$.

\begin{figure}
\includegraphics[width=8.5cm]{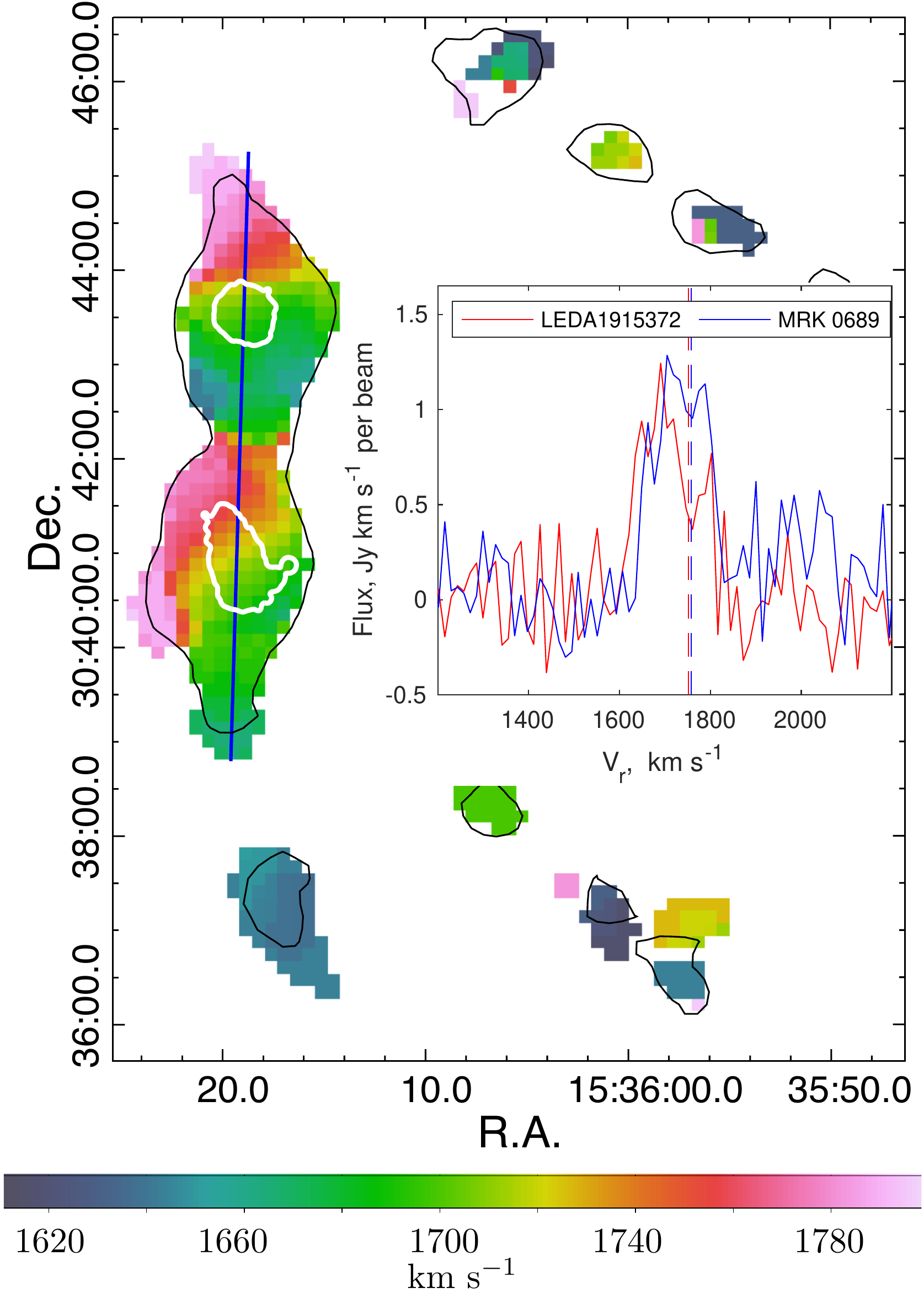}
\caption{ The \hi\ velocity field from the GMRT low-resolution cube. The black contour traces the lowest column density contour of the low-resolution integrated \hi\ image (N(\hi) = 0.5 $\times$ 10$^{20}$ cm$^{-2}$). The white contours trace the optical $g$-band light at the surface brightness levels of 26 mag arsec$^{-2}$. The solid blue line represents the position of the P$-$V slice (see Figure \ref{pvdiagram}). The inset shows the integrated \hi\ spectrum of  LEDA1915372 and MRK 0689. The vertical dashed line represents their respective optical velocities.}
\label{vmap}
\end{figure}

\begin{figure}
\includegraphics[width=8.5cm]{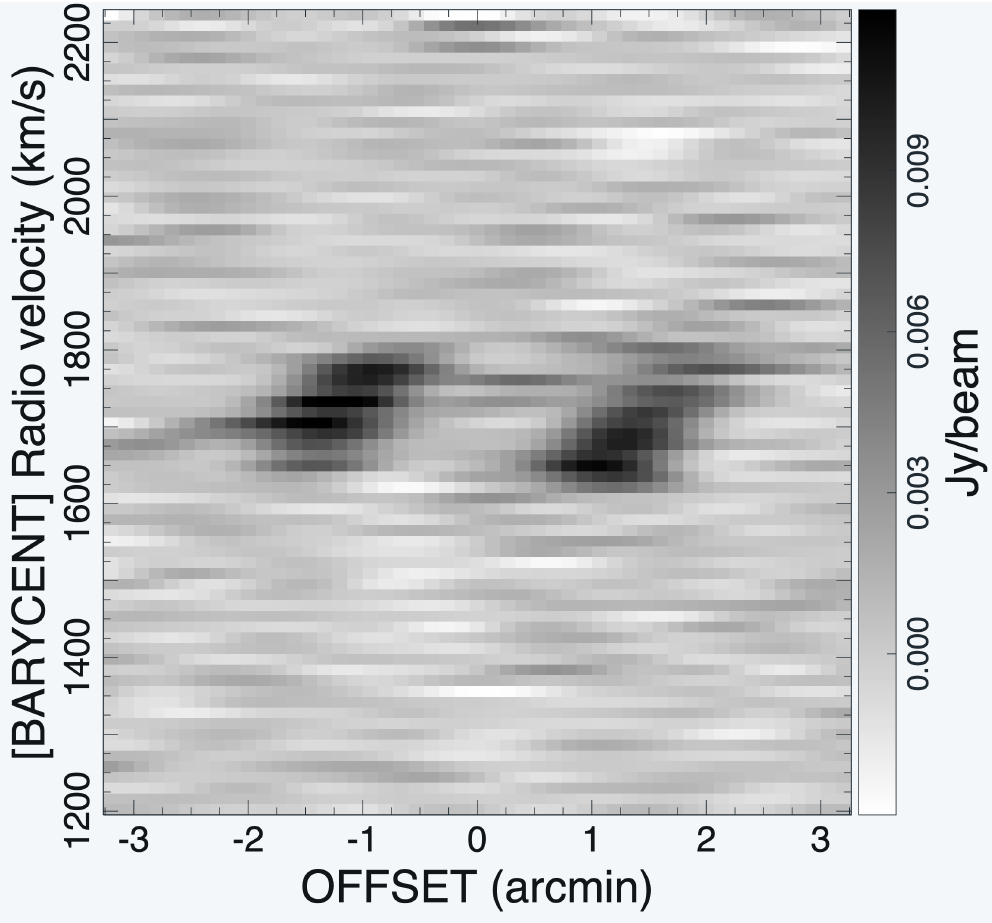}
\caption{ P-V diagram along a line crossing the \hi\ bridge and the two galaxies (see Figure \ref{vmap}) with a width of the slice of 56$\arcsec$. 
We produced the P$-$V diagram using CARTA v3.0.0. }
\label{pvdiagram}
\end{figure}

Figure \ref{himap} shows the contours obtained from the integrated \hi\ map overlaid on the Legacy $g$, $r$ and $z$-band combined tricolor image. 
The green contour represents the low-resolution \hi\ cube. 
The field view of the image is 12$\arcmin$$\times$12$\arcmin$.  
The \hi\ bridge is clearly visible, validating an interaction between LEDA\,1915372 and MRK\,0689. 
We show the beam size in the upper right corner, which is significantly larger than the galaxy extension. 
The lowest contour corresponds to 5$\times$10$^{19}$ cm$^{-2}$ ($\sim$5$\sigma$ assuming a 20 \kms\ line width). 
Looking at the detached contours around the galaxies, it appears that some areas of extended diffuse emission are separate from the main body of the galaxies. However, it is important to note that the identification of these detached diffuse regions is only marginally above the signal-to-noise ratio (SNR) threshold in the current dataset.  Consequently, we have excluded these faint detections from any further analysis.

Further, our data cube is also processed using the powerful SoFiA-2 source-finding algorithm. By employing the default parameters of SoFiA-2, we utilized the S+C source-finding algorithm, incorporating a threshold of 4$\sigma$ for optimal sensitivity and reliability. The reliability parameter was set to ``true" while maintaining the default parameters. This comprehensive analysis successfully verifies the existence of the HI bridge connecting LEDA\,1915372 and MRK\,0689.

The insets of Figure \ref{himap} illustrate the contour maps obtained from high-resolution synthesized beams, overlayed on the $g-z$ color maps. 
In LEDA 1915372, the \hi\ gas is noticeably displaced to the southwest with a tail that points northward.
Furthermore, the \hi\ emission peak is co-located with Blue Plume, indicating its tidal origin. 
The high-resolution \hi\ map of MRK\,0689 also shows the extension of diffuse emission toward LEDA\,1915372.

The \hi\, tidal tail structure and kinematics are further illustrated in the channel maps shown in Figure \ref{chmap}. To guide the position of the channel map, we also show a grayscale image of the system in the background where the green contours represent the lowest column density level shown in Figure \ref{himap}.  The connection in \hi\, between interacting galaxies becomes clearer in the channel maps of velocity range  1635$-$1789 \kms.

Figure \ref{vmap} shows the velocity field of the system. 
In both LEDA\,1915372 and MRK\,0689, we detect a pronounced velocity gradient. 
The overall gradient is primarily aligned with the major axis of the main stellar body of MRK\,0689, implying a shared direction of angular momentum. 
Additionally, we provide an inset displaying the integrated spectrum of both galaxies.
Notably, the middle dip of the \hi\ velocity profile accurately corresponds with their respective optical velocities.

Figure \ref{pvdiagram} shows a position-velocity (P$-$V) diagram of the system. The P$-$V diagram was produced using CARTA v3.0\footnote{https://cartavis.org/}, by crossing a line region over the \hi\ bridge and the two galaxies. The S/N is poor in the P$-$V diagram, particularly in the bridge area. However, it successfully shows the presence of gradient and connection between the two galaxies.

The integrated \hi\ emission-line flux of the system was measured using a large 9$\arcmin$ diameter aperture centering at  MRK\,0689, yielding a total flux of 9.64 Jy \kms. 
The value is slightly lower than the 21 cm emission-line flux reported in NED, which was measured by a 48 m single-dish Green Bank Telescope Telescope (GBT) with a 9$\arcmin$ beam in the L band. The ALFALFA survey, conducted using the Arecibo Telescope with a 3$\arcmin$.5 beam size, detected MRK\,0689. The cataloged value of the 21 cm emission-line flux for MRK\,0689 (AGC\,251077) in the ALFALFA survey was 7.16 Jy km$^{-1}$. To corroborate this finding, we measured the flux of MRK\,0689 using a similar 3$\arcmin$.5 diameter aperture and obtained a value of 6.31 Jy km$^{-1}$. This small discrepancy could be attributed to factors such as loss of UV coverage due to flagging.  Also, the GMRT synthesized beam could have resolved diffuse emission from the extended and low column density regions. 
However, by using a much larger aperture of 12$\arcmin$ that covers the entire area of both interacting galaxies and detached diffused regions, we obtained a total emission flux of 13.05 Jy \kms. 
The integrated \hi\ line flux density of the LEDA\,1915372 and MRK\,0689, estimated from high-resolution maps, are 5.01 and 5.94 Jy \kms, respectively.

\section{Discussion and Conclusion}

This study investigates an interacting system of two dwarf galaxies exhibiting distinct morphologies. The optical and radio 21 cm emission-line observations provide clear evidence of an ongoing interaction between the two galaxies. 
In the subsequent sections, we discuss potential scenarios for the origin of \hi\ in dEs and the nature of the central star formation.

\subsection{Origin of \hi\ in dEs}

Observations have revealed the presence of \hi\ gas within giant and normal elliptical galaxies located in field environments. 
Early-type galaxies are believed to acquire gas via various forms of accretion, such as the absorption of gas-rich satellites, cooling of ionized gas, and cold-mode accretion from the intergalactic medium.
The presence of recent or ongoing accretion onto galaxies, as highlighted by \cite{Oosterloo10}, usually results in a surge of star formation. Conversely, despite possessing a high gas content, inactive accretion tends to lead to a dearth of star formation. However, the origin of \hi\ in dEs remains poorly understood due to a lack of adequate studies in both theoretical and observational domains.

Using optical and 21 cm radio data, we have observed an ongoing interaction between LEDA\,1915372 and MRK\,0689. 
This interaction unquestionably plays a role in the revival of star formation activity in LEDA\,1915372. 
However, the crucial question that remains is whether the \hi\ gas present in LEDA\,1915372 is intrinsic to the galaxy or  if it was acquired from MRK\,0689. 
Our assessment of the \hi\ emission line indicates that both galaxies have a comparable amount of \hi\ mass. 
This raises the possibility that a significant transfer of gas mass may have occurred between the two galaxies in the past. This could imply that almost half of the gas mass was transferred from MRK\,0689 to the neighboring companion, LEDA\,1915372.

One intriguing aspect of LEDA 1915372 is that most of its gas mass is located outside the galaxy's main stellar body and instead is displaced around the Blue Plume, a faint tidal feature.
Notably, the Blue Plume harbors a relatively young stellar population, suggesting the tidal feature is built up through newly formed stars. 
Such a blue tidal feature is rare in observations, especially in dEs. 
In contrast, the \hi\ distribution in MRK\,0689 is colocated with its main stellar body. 
The optical color map shows that the extended tidal feature of MRK\,0689 is redder than its main body, providing an interesting contrast to the blue tidal feature observed in LEDA 1915372.

The \hi\ kinematics of both galaxies exhibit similar rotation directions, indicating that they share angular momentum. 
Moreover, both galaxies have nearly identical emission-line metallicities (12+log[O/H]), suggesting that they have undergone similar gas enrichment histories. This provides compelling evidence for the shared origin of \hi\ in both galaxies. 
If LEDA\,1915372 were made up of recycled or leftover gas from aging galaxies, the emission-line metallicity in dEs would typically be higher than what we currently observe in LEDA\,1915372, as suggested by \citet{Paudel20}.

Assuming that the Blue Plume was entirely created by a tidal interaction, its $g-r$ color of 0.34 mag suggests that the interaction age is at least 1 Giga year or more \citep{Urich17}. 
It is likely that LEDA\,1915372 accreted gas during an interaction with MRK\,0689 long ago, which fueled star formation at the center of LEDA\,1915372 and formed a blue tidal tail. 
Determining the exact age of the interacting system is beyond the scope of this work as it requires a detailed analysis of the stellar population properties using spatially resolved optical spectroscopic data.

 \subsection{Off-centered star formation in dEs}
 
In our previous study \citep{Paudel20}, we reported the presence of the central starburst region in dE galaxies, which could potentially evolve into young nuclei. 
The central, compact, blue star-forming blob is often positioned slightly off-center from the dE center, as in the case of LEDA\,1915372.  
This raises the question of whether the off-centered starburst activity is a sign of external accretion.

Off-centered nuclear starburst activity can be triggered by the gravitational force from an imperfectly aligned merger. 
High-resolution simulations have shown that interaction-driven compression is most likely to trigger starburst activity in the outer regions of galaxies \citep{Renaud22}, which is consistent with the off-centered starbursts observed in many interacting systems \citep{Jarrett06}. This is especially important when a galaxy accretes a lump of \hi\ gas with a mass significantly less than the galaxy itself. As the host and the gas lumps come closer, the gravitational force compresses the gas before it reaches the center. 
On the other hand, the high angular momentum in the gas prohibits radial inflow to the center, and a starburst episode can be ignited in the compressed gas mass before it falls to the galaxy's center \citep{Blumenthal18}. 
In this context, understanding the dynamics of off-centered nuclear starbursts in interacting galaxies can help to explain how off-centered nuclei are formed in dwarf galaxies.

\newpage
\begin{acknowledgments}
S.-J.Y. and S.P. acknowledge support from the Basic Science Research Program (2022R1A6A1A03053472) through the National Research Foundation (NRF) of Korea. S.P. and S.-J.Y., respectively, acknowledge support from the Mid-career Researcher Program (No. RS-2023-00208957) and the Mid-career Researcher Program (No. 2019R1A2C3006242) through the NRF of Korea.
O.B. is supported by the {\em AstroSignals} Sinergia Project funded by the Swiss National Science Foundation.

We thank the staff of the GMRT that made these observations possible. GMRT is run by the National Centre for Radio Astrophysics of the Tata Institute of Fundamental Research.

This study uses archival images from the Legacy survey. The DESI Legacy Imaging Surveys consist of three individual and complementary projects: the Dark Energy Camera Legacy Survey (DECaLS), the Beijing-Arizona Sky Survey (BASS), and the Mayall z-band Legacy Survey (MzLS). DECaLS, BASS and MzLS together include data obtained, respectively, at the Blanco telescope, Cerro Tololo Inter-American Observatory, NSF’s NOIRLab; the Bok telescope, Steward Observatory, University of Arizona; and the Mayall telescope, Kitt Peak National Observatory, NOIRLab. NOIRLab is operated by the Association of Universities for Research in Astronomy (AURA) under a cooperative agreement with the National Science Foundation. Pipeline processing and analyses of the data were supported by NOIRLab and the Lawrence Berkeley National Laboratory (LBNL). Legacy Surveys also uses data products from the Near-Earth Object Wide-field Infrared Survey Explorer (NEOWISE), a project of the Jet Propulsion Laboratory/California Institute of Technology, funded by the National Aeronautics and Space Administration. Legacy Surveys was supported by: the Director, Office of Science, Office of High Energy Physics of the U.S. Department of Energy; the National Energy Research Scientific Computing Center, a DOE Office of Science User Facility; the U.S. National Science Foundation, Division of Astronomical Sciences; the National Astronomical Observatories of China, the Chinese Academy of Sciences and the Chinese National Natural Science Foundation. LBNL is managed by the Regents of the University of California under contract to the U.S. Department of Energy. The complete acknowledgments can be found at https://www.legacysurvey.org/acknowledgment/.

\end{acknowledgments}


\end{document}